\documentclass[12pt]{article}

\usepackage{microtype}
\usepackage{graphicx}
\usepackage{amsmath, amsthm, amssymb}
\usepackage{bm}
\usepackage{natbib}
\usepackage[unicode, colorlinks]{hyperref}
\usepackage{enumitem}
\usepackage{booktabs}
\usepackage{multirow}
\usepackage{tabularx}
\usepackage[dvipsnames]{xcolor}
\usepackage[skip=1em]{parskip}
\usepackage{ulem}

\usepackage[document]{ragged2e}      
\usepackage{setspace}                
\usepackage{newtxtext,newtxmath}     
\usepackage{geometry}
\usepackage{tikz}
\usepackage{tikzscale}
\usetikzlibrary{arrows, backgrounds, patterns, matrix, shapes, fit,
  calc, shadows, plotmarks, babel, positioning}

\newcommand{\eat}[1]{}

\graphicspath{{./figures/}}

\theoremstyle{remark}

\hypersetup{%
  linkcolor=NavyBlue,%
  citecolor=NavyBlue,%
  filecolor=Red,%
  urlcolor=NavyBlue,%
  menucolor=Red,%
  runcolor=Red,%
  linkbordercolor=NavyBlue,%
  citebordercolor=NavyBlue,%
  filebordercolor=Red,%
  urlbordercolor=NavyBlue,%
  menubordercolor=Red,%
  runbordercolor=Red
}

\hypersetup{%
  pdftitle={Bayesian Graphical Entity Resolution using Exchangeable Random Partition Priors},%
  pdfauthor={Anonymous Authors},%
  pdfkeywords={entity resolution, record linkage, Bayesian models, exchangeability, random partitions, Ewens-Pitman random partitions}%
}

\makeatletter
\newcommand\sosname{Statement of Significance}
\if@titlepage
  \newenvironment{sos}{%
      \titlepage
      \null\vfil
      \@beginparpenalty\@lowpenalty
      \begin{center}%
        \bfseries \sosname
        \@endparpenalty\@M
      \end{center}}%
     {\par\vfil\null\endtitlepage}
\else
  
\fi

\makeatother

\title{A Primer on the Data Cleaning Pipeline}
\author{}
\date{}

\begin{document}

\makeatletter
\begin{titlepage}
    \null
    \vskip 3em%
    \begin{center}%
        {\LARGE \@title \par}%
        \vskip 3em%
        {\large 
            \lineskip .5em%
            Rebecca C.~Steorts$^{*}$ \\[1em]
            $^{*}$ Rebecca C. Steorts is an Associate Professor in the Department of Statistical Science at Duke University with affiliations in Computer Science, Biostatistics and Bioinformatics, the Social Science initiate at Duke, and the Rhodes Information Initiative at Duke. Steorts is a Mathematical Statistician at the United States Census Bureau. \par}%
        \vskip 3em%
    \end{center}
    {\setlength{\extrarowheight}{14pt}%
    \begin{tabularx}{\textwidth}{@{}>{\bfseries}l X}%
        Corresponding author: & 
            Rebecca C. Steorts \newline
           Old Chemistry Hall \newline
           Durham, North Carolina 27708 \newline
            Email: \hyperref{mailto:beka@stat.duke.edu}{}{}{beka@stat.duke.edu, rebecca.carter.steorts@census.gov} \\
        Word count: &  4048 words excluding captions and references  \\
        Funding sources: & Steorts is supported by an NSF CAREER Award. 
    \end{tabularx}\par}%
    \par
    \vskip 1.5em
\end{titlepage}
\makeatother

\clearpage

\maketitle
\doublespacing
\begin{abstract}
The availability of both structured and unstructured databases, such as electronic health data, social media data, patent data, and surveys that are often updated in real time, among others, has grown rapidly over the past decade. With this expansion, the statistical and methodological questions around data integration, or rather merging multiple data sources, has also grown. Specifically, the science of the ``data cleaning pipeline'' contains four stages that allow an analyst to perform downstream tasks, predictive analyses, or statistical analyses on ``cleaned data.'' 
This article provides a review of this emerging field, introducing technical terminology and commonly used methods. 
\end{abstract}

\textbf{Statement of Significance}: The article reviews the data cleaning pipeline, introducing technical terminology and commonly used methods.

\newpage
\section{Introduction}
\label{sec:data-cleaning-pipeline}

The availability of databases, such as electronic health data, social media data, patent databases, administrative and commercial data, geodata, digital trace data, and surveys that are often updated in real time, among others, has grown rapidly over the past decade. This expansion of data has generated widespread interest in the statistical, methodological, and ethical issues surrounding the integration (or merging) of multiple data sources \citep{fienberg2009integrated, kaplan2022practical, sadinle2018bayesian, aleshin2022multifile, wick2009entity, culotta2007canonicalization, binette2022reliability}. Application areas of interest include statistical science, computer science, engineering, and machine learning \citep{Herzog_2007, christen_data_2012, Winkler2014, Jurek2019, Christophides2019, asher2020introduction, Papadakis2021, binette2022almost}, the social sciences \citep{mcveigh2019scaling, enamorado2019using}, health care \citep{shan2022bayesian, Rogot1986, meray2007probabilistic,  Jaro1995, Gutman2013, shan2020bayesian, farley2020bayesian}, official statistics and survey methodology \citep{Jaro1989, Winkler1990, Fortinietal01, Chevrette2011GLINKA, Dasylva2014, Dasylva2018}, human rights statistics \citep{lum2013applications, price2015documents, Sadosky2015, sadinle_detecting_2014, sadinle2018bayesian, aleshin2022multifile}, author name disambiguation \citep{lai_2011, louppe2016ethnicity, zhang2018name, subramanian2021s2and, liu2021oag}, and forensic science \citep{Tai2019, tai2020automatically}.

\begin{figure}[htbp]
\begin{center}
\includegraphics[width=1\textwidth]{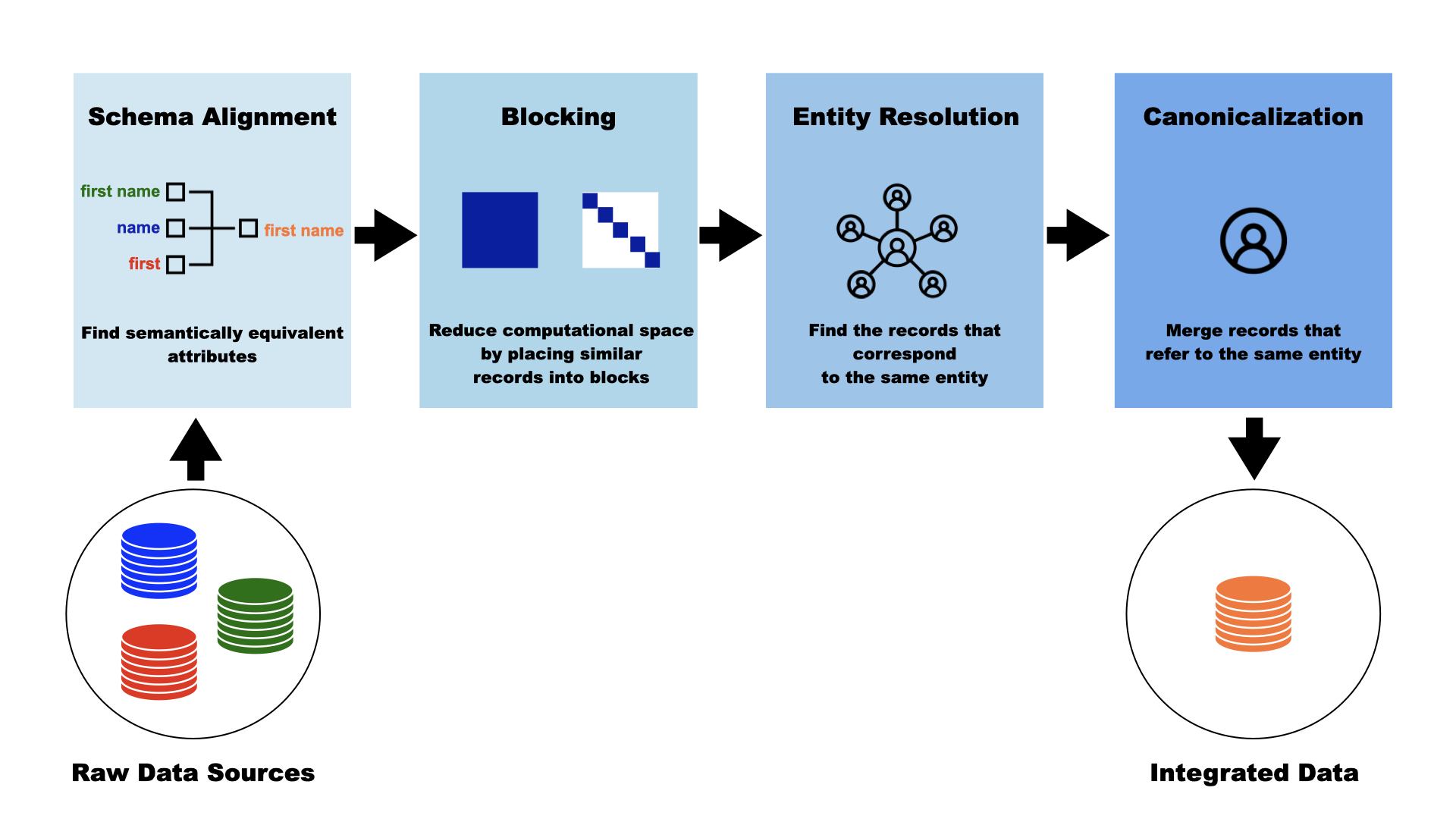}
\caption{We illustrate the four stages of data cleaning pipeline: 1) schema alignment, 2) blocking, 3) entity resolution, and 4) canonicalization. Observe that the input to the schema alignment stage is multiple, raw data sources. The output after canonicalization is a single integrated database with unique identifiers for usage in downstream tasks and statistical analyses.}
\label{fig:data-clean}
\end{center}
\end{figure}

The main goal of merging databases is answering statistical analyses (downstream tasks) on ``cleaned data." As such, one must perform the four stages of the ``data cleaning pipeline'' to have more reliable conclusions. Figure \ref{fig:data-clean} provides a visual representation of the pipeline. For simplicity, we consider structured data, such as name, address, and zip code. We omit unstructured data, such as paragraphs or images, from our review. For more information regarding how to handle both structured and unstructured data, we refer to Papadakis et al. (2022).

The first stage of the data cleaning pipeline, known as \textit{schema alignment}, takes multiple data sources and forces the attributes (fields or features) to be \textit{structured and aligned}. For example, multiple data sources contain raw, messy data that needs to be converted such that the rows correspond to records and each column corresponds to an attribute. 

In the second stage, the schema aligned data undergo \textit{blocking}, which places similar records into ``bins,'' called \textit{blocks}. The goal of blocking is to avoid all-to-all record comparisons, which scales quadratically (with the number of records). 

In the third stage, the blocked data undergo \textit{entity resolution} (record linkage or de-duplication), which removes duplications within the blocks. The output of the entity resolution stage is clusters of records that correspond to the same entity.

In the fourth and final stage, each cluster of records that corresponds to a single entity undergoes \textit{canonicalization}, \textit{data fusion}, or \textit{merging}, which identifies the most representative record in the cluster. Said differently, the canonicalization stage attempts to repair any errors from previous stages to output a single integrated data set, with a unique identifier for each record. Each stage builds upon the previous one; any errors made in a previous stage propagate into the next stage. 
Finally, the output from canonicalization is for a \textit{downstream task}, which is a prediction, inference, or statistical analysis. 

The main goal of the data cleaning pipeline is to be able to provide meaningful statistical analyses on the final integrated data source. Common examples of downstream tasks include multiple systems estimation (capture recapture analysis or population size estimation), small area estimation, composite estimation, or model-building exercises, such as regression.
The remainder of the article focuses on reviewing the four stages of the data cleaning pipeline. Also, we review common approaches that are utilized after the data cleaning pipeline, where statistical analyses or downstream tasks are performed. We conclude with a summary.

\section{A Review of the Data Cleaning Pipeline}
\label{sec:review-data-clean}
In this section, we introduce and review the four stages of the data cleaning pipeline that allow one to perform downstream tasks and statistical analyses.

\subsection{A Review of Schema Alignment}
\label{sec:schema-alignment}

In this section, we review terminology and present simple approaches.

\subsubsection{Terminology and Background}

A \textit{schema} (or relational database), which derives from the computer science literature, is an organizational structure for a database. In the context of this problem, a schema consists of all attributes for the input data sources. Disparate data sources may use different naming conventions for records that refer to the same attribute. For instance, the schema alignment stage attempts to identify that the attributes ``first name,'' ``name,''  and ``first'' all refer to a global (or common) attribute ``first name.'' This stage replaces the aforementioned attributes with the global attribute ``first name.'' Therefore, \textit{schema alignment} is essential when the multiple input data sources lack attribute agreement. 

More formally, this stage attempts to find ``semantically equivalent attributes'' \citep{Papadakis2021} and/or leverage attribute information. For example, attribute mappings between the data sources provide information regarding ``transformations, mappings, or rules'' \citep{Papadakis2021}. For instance, it is beneficial to learn mappings or correspondences between acronyms of attributes, such as ``North Carolina" versus ``NC." Other examples include: ``Street" versus "St." or "first" versus "1st."

Methods using a schema are \textit{schema-based} or \textit{schema-aware}. 
On the other hand, data sources that lack or ignore this information are \textit{schema agnostic} or \textit{schema-independent}.
In many applications, an analyst may receive data that has already been schema aligned, and thus, would lack information regarding the original raw data sources. In these cases, one would not need to perform this stage.

\subsubsection{Approaches for Schema Alignment}

\begin{figure}[htbp]
\begin{center}
\includegraphics[width=\textwidth]{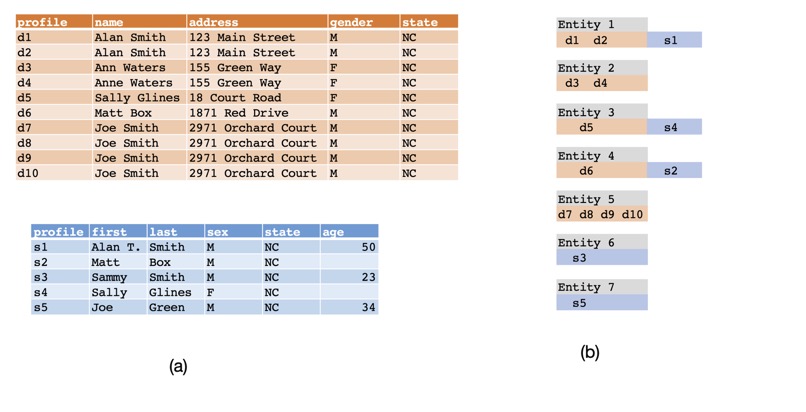}
\caption{(a): Two disparate schema. (b): The corresponding profiles (or records) that correspond to each unique entity.}
\label{fig:schema}
\end{center}
\end{figure}

Schema alignment requires the analyst to identify all attributes that lack agreement and identify a common naming convention. Next, the analyst updates the common naming convention, such as manually. While this approach is simple, it is not reproducible or easily documented. Importantly, given its manual nature, mistakes can easily occur. Figure \ref{fig:schema} provides two disparate schema needing attribute alignment. We create a mapping between first/last and name. In addition, we map gender and sex. 

As already mentioned, some methods learn attribute mappings between the data sources, which can provide information regarding “transformations, mappings, or rules" of the attributes \citep{winkler2001record, tejada2002learning}. Both \cite{winkler2001record} and \cite{tejada2002learning} recommend rule-based systems, where the rules can differ depending on the linked data sources. For example, a simple rule may require attribute alignment if the attributes are semantically equivalent. A concrete example of such a rule is to recognize that ``first name," ``name," and ``first" all identify a person's first name. While rules or more automated methods can be utilized, they will always rely on manual review of an analyst. No automated system is perfect; it may be difficult to find errors for complex rules.

Many proposed methods for performing schema alignment exist within the computer science, machine learning, and database management literatures. 
As already mentioned, we only consider structured data. In the case of both structured and unstructured data, the first stage requires \textit{schema refinement} (instead of alignment). We refer to \cite{Papadakis2021} for an introduction to this extension.

\subsection{A Review of Blocking}
\label{sec:blocking}
In this section, we briefly review blocking.
Blocking places similar records into ``bins," providing a dimensionality reduction. The blocks can be overlapping or non-overlapping. Note that blocking does not refer to other definitions in other areas, such as in clinical trials or analysis of variance.

Typically, blocking operates (internally) in a schema-aware manner. As such, this step assumes that the data sources are schema`aligned. Thus, analysts must decide the most useful attributes that will lead to representative signatures, known as \textit{blocking keys}. Users place similar or identical blocking keys into the same partition or bin \citep{christen_2012, steorts_comparison_2014, murray2016probabilistic, Papadakis2021}. 

Formally, given data sources $d_1, \ldots, d_k$ a blocking method (with a key) $B^{\text{key}}$ generates a set of blocks 
$\{b_1^{\text{key}}, \ldots, b_{n}^{\text{key}} \}.$ Each blocking key $b_{i}^{\text{key}}$ is a subset of records (or profiles) from the data sources that have the same (or similar) value for key. That is, $b_{i}^{\text{key}} \subset  \cup_{j=1}^k d_j.$  \cite{Papadakis2021} provides illustrations regarding blocking keys.

The simplest method for constructing blocks rely on attributes (e.g., gender or zip code). \textit{Traditional blocking} methods place records in the same block if and only if they agree on the blocking attributes. In addition, traditional blocking, while computationally efficient, makes an unrealistic assumption that the blocking keys are error free \citep{steorts_comparison_2014, Papadakis2021}.
We can form more complex rules based upon the attributes that involve unions and intersections of the attributes, known as \textit{conjunctions} (or disjunctions of conjunctions). 
These methods require an expert to define a transformation of one or more attributes as the blocking key of each record that occurs in a data source (or profile). 

\begin{figure}[htbp]
\begin{center}
\label{fig:block}
\includegraphics[width=\textwidth]{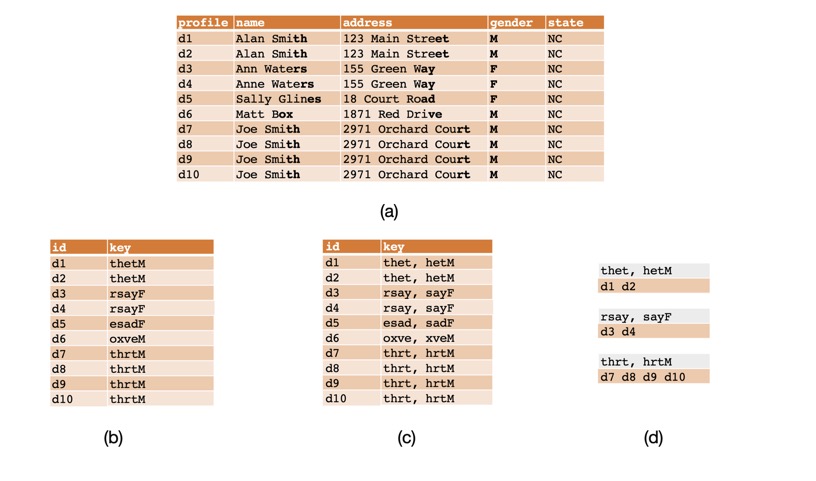}
\caption{(a): Toy illustration of a data source with the standard blocking criterion bolded. (b): Standard blocking keys. (c): 4-grams based upon the standard blocking in (b). (d): The blocks from 4-grams, which are the same as for standard blocking with the exception that the blocking keys are different.}
\end{center}
\end{figure}

On the other hand, \textit{probabilistic approaches} do not assume part of the attributes are fixed and error-free. Instead, these approaches utilize the entire records, resulting in a more data-driven approach. This involves replacing each record by a multi-set of length $k$ contiguous substrings, called \textit{k-grams}, \textit{k-shingles}, or \textit{k-tokens}.
Figure \ref{fig:block} provides a simple illustration of one data source, where we consider two blocking criteria. See \cite{steorts_comparison_2014, Papadakis2021} for more advanced probabilistic blocking.

One of the most widely used blocking methods is locality sensitive hashing (LSH), which is a probabilistic method for dimension reduction that has strong theoretical properties. Specifically, LSH is a quick way of finding approximate nearest neighbors, which is one way of creating blocks. Many variants of LSH for blocking and are discussed in the previous references. Finally, supervised, unsupervised, and hybrid approaches are appropriate for blocking; however, this is not optimal procedure for all applications \citep{christen_2012, Papadakis2021}.

\subsection{A Review of Entity Resolution}
\label{sec-review-er}
In this section, we review comparison-based methods (Fellegi-Sunter and supervised learning methods) and unsupervised learning methods (comparison- and direct-based methods).

\subsubsection{Fellegi-Sunter Based Methods}
\label{sec:fs}

The earliest published work is that of \cite{Dunn1946} who believed that it was important to document record keeping for official statistical agencies.
Roughly ten years later, \cite{newcombe_automatic_1959} proposed one of the first algorithms for entity resolution. \citet{fellegi_theory_1969} formalized the work of \cite{newcombe_automatic_1959} via a likelihood ratio test based upon agreement patterns of record pairs and classified the records as \textit{links}, \textit{non-links}, or \textit{possible-links}. Their method --- the Fellegi-Sunter (FS) method --- was shown to be optimal under statistical decision theory. The authors provided two variants of their method, one exact and one approximate via the expectation–maximization algorithm, where the former is more scalable and practical. The FS method is one of the most popular methods utilized in statistical entity resolution (and other fields) due to its simplicity, statistical optimality, and scalability. Moreover, researchers have extended it from both methodological and computational viewpoints  \citep{winkler_overview_2006, 
sadinle_generalized_2013, enamorado_using_2019,  mcveigh2019scaling, sadinle_detecting_2014, sadinle2018bayesian, aleshin2022multifile, tang2020, Armstrong1992, Winkler1992, Winkler1993, Thibaudeau1993, Smith1975, belin1990proposed, Armstrong1992, belin_1995, Nigam2000, larsen_2001, lahiri_2005,wortman2019record}.

\subsubsection{Supervised Entity Resolution}
\label{sec:supervised}

A supervised learning method requires a comparison vector consisting of record pairs --- the same input as required as the FS method. Many packages have implemented supervised learning methods that require a binary comparison vector. More specifically, supervised learning methods require training data and testing data. The training data may come from manually labeled data of record pairs in the form of match or non-match. The user builds a classifier (using the training data) that predicts the match or non-match status (on the testing data). 
The resulting output is the match and non-match status for all record pairs. 

Supervised entity resolution employs training data to develop classification methods
\citep{Trajtenberg2008, Azoulay2011, Bailey2017} or crowdsourcing \citep{Sarawagi2002, Wang2012, Vesdapunt2014, gokhale2014corleone, frisoli2019novel}. 
Fully supervised methods utilize large amounts of training data, which contrasts semi-supervised methods, which often assume a minimal 
amount of training data. Examples of supervised models include logistic regression, random forests, support vector machines, Bayesian additive regression trees, among other types of methods that originated in the machine learning literature \citep{hastie_2001}. Neural networks are popular for supervised entity resolution \citep{Gottapu2016, Ebraheem2017, Ebraheem2018, Mudgal2018, Kooli2018, Kasai2020, barlaug2021neural}. For more information on supervised learning methods, we refer to \cite{Papadakis2021}.

The aforementioned methods are useful when the training data is extremely accurate. Furthermore, these methods scale to extremely large databases.  Supervised learning methods are sensitive to tuning parameters and the amount of training data needed, which can be seen using standard evaluation metrics used in the literature such as recall (true positive rate or sensitivity) and precision (positive predictive value). In addition, using too much training data can lead to biases in predictions or inferences downstream. Finally, most classification methods do not quantify uncertainty of the entity resolution task into later statistical analyses.

\subsubsection{Unsupervised Entity Resolution}

In this section, we review the two main approaches in the literature to unsupervised (or Bayesian) entity resolution.  
First, we review Bayesian FS methods, which extend the seminal work of \cite{fellegi_theory_1969} to an unsupervised setting. This framework traditionally models the comparisons of the attributes between pairs of records through a mixture model \citep{larsen_2001, Jaro1989}. This is known as the \textit{comparison-based framework} \citep{sadinle_detecting_2014, sadinle2017bayesian, sadinle2018bayesian, aleshin2022multifile}. 

Second, we review unsupervised entity resolution methods, where the goal is to cluster records to an unknown, latent entity. These methods directly model the attribute-level distortion, noise, and/or transcription errors of a record \citep{Matsakis10, liseo_jos_2011, liseo_2011, liseo_2013, gutman2013bayesian, steorts14smered, tancredi_2015_regression, steorts_entity_2015, steorts_bayesian_2016, steorts2018generalized, DalzellReiter18, tancredi2018unified, steorts_performance_2017, tang2020, kaplan2018posterior, marchant2021d}. This is known as  the \textit{direct-modeling framework}.

\paragraph{Comparison-Based Methods}

Comparison-based methods of record pairs became popular with the framework of \cite{fellegi_theory_1969}. This work led to many extensions previously mentioned. In this section, we review unsupervised (or Bayesian) extensions. \cite{sadinle_detecting_2014} extended \cite{fellegi_theory_1969}, considering the case of \textit{duplicate detection} or \textit{de-duplication}, or the process of removing duplicate records within a single database.

One extension made was providing transitive closures regarding the match status. (Note that a linkage is transitive in the following sense: if records A and B are the same individual, and records B and C are the same individual, then records A and C must be the same individual as well.)
The coreference matrix is a ``partition of the data file that groups co-referent records together'' \citep{sadinle_detecting_2014}. In addition, \cite{sadinle_detecting_2014} included prior information, if available, on the match and non-match probabilities. This provided a posterior distribution on the co-reference matrix (or linkage structure), which provides a partition for representing linkages of records across and within databases. For example, assume two databases with record $i$ in the first database and record $j$ in the second database. More formally, define the entries of the co-reference matrix such that $c_{i,j} = 1$ if records $i$ and $j$ are the same record. Otherwise, $c_{i,j} = 0.$ Furthermore, the proposed Bayesian methodology allows for uncertainty quantification regarding the match and non-match status. 

\cite{sadinle2017bayesian} considered the setting of \textit{bipartite record linkage}, where the goal is to merge two databases. Duplications could occur across databases but not within databases. This assumption imposed is called a \textit{one-to-one restriction} regarding the linkage across the two databases. In short, a record in the first database can match a maximum of one time to a record in the second database. \cite{fellegi_theory_1969} (and most extensions) ignored this assumption.  \cite{sadinle2017bayesian} proposed incorporating missing values handling partial agreements. To our knowledge, \cite{sadinle2017bayesian} provided the first extensions using generalized loss functions that allow for a rejection option, providing an alternative to the traditional FS decision rule. 

\cite{sadinle2018bayesian} extended the bipartite record linkage framework to include the case of entity resolution, called joint duplicate detection and record linkage. This provided a general framework for duplications within and across all databases. \textit{Linkage averaging} was introduced and propagates the uncertainty of the record linkage method to a downstream task, such as multiple systems estimation. The proposed methodology is illustrated on a case study for human rights violations in El Salvador.

\cite{aleshin2022multifile} considered Bayesian entity resolution models, where a structured prior is proposed on the coreference matrix. For more information, we refer to details in the paper regarding the construction of the prior distribution. Similar to earlier work, \cite{aleshin2022multifile} derived loss functions with an abstain option and illustrated their proposed methodology on real data and simulation studies.

\paragraph{Direct-Modeling Based Methods}

Other approaches assume a model at the attribute-level of each record. These models cluster observed records (at the attribute-level) to a latent (unknown) entity, which avoids all-to-all record comparisons. A \textit{linkage structure} partitions co-referent (or duplicated) records into the same group. (The linkage structure is also a co-reference matrix.) A user must define a likelihood for the observed records (at the attribute-level) in the data sources. One must place priors on any unknown parameters, such as the linkage structure. 
Such a specification provides for a posterior distribution on the linkage structure \citep{Matsakis10, liseo_jos_2011, liseo_2011, liseo_2013, gutman2013bayesian, steorts14smered, steorts_entity_2015, steorts_bayesian_2016}.

Many models assume the records contain distortions, noise, and transcription errors. As such, the likelihood function handles these ``distortions."
One way to model the distortion of the records is to use a latent variable model, where the distortion is a latent (or unknown, random) variable. 
For example, \cite{liseo_2011} adapted the hit and miss model of \cite{copas_record_1990} to entity resolution, proposing a measurement error model for distortions.

In addition to the assumptions above, some researchers proposed that the attributes were only categorical-valued \citep{Matsakis10, liseo_2011, steorts14smered, steorts_bayesian_2016}. On the other hand, \cite{liseo_2011} proposed continuous attributes under a normality assumption. \cite{steorts_entity_2015} extended this prior work to include both categorical and string-based attributes. \cite{liseo_2011, steorts14smered, steorts_bayesian_2016} provided for Bayesian decision rules under a given loss function. This approach allows for uncertainty quantification regarding the linkage structure (or match and non-match status) and functions of the linkage structure. One major limitation of both direct- and comparison-based methods has been the ability to scale to moderate to large databases. 

\cite{marchant2021d} proposed one way to scale to larger databases containing over a million total records. Specifically, they proposed a joint blocking and entity resolution framework, extending the work of \cite{steorts_entity_2015}. They proposed an \textit{auxiliary variable representation} of \cite{steorts_entity_2015} which partitions the latent entities and records into blocks. The work of \cite{marchant2021d} differs from typical blocking in the literature as the assignments of latent entities and records to blocks are \textit{random} and \textit{jointly inferred} with other parameters of the model.  Furthermore, they formally proved that the auxiliary variable representation preserves the marginal posterior distribution over the model parameters. This result implies that any inferences are independent of the blocking criteria. They utilized a distributed partially-collapsed Gibbs sampler to perform inference and other computational optimizations, where they investigated their model on many data sets as well as a case study from the United States Census Bureau.

This prior work has led to interesting discoveries both in clustering and in entity resolution. The \textit{microclustering property} describes the sub-linear growth of entity resolution (and other clustering tasks, such as community detection). The size of the clusters grow sub-linearly as the total number of records grow \citep{zanella_flexible_2016, betancourt2022random, betancourt2022prior}. Therefore, applying a Bayesian nonparametric (BNP) model with a large number of data points (records) might tend to favor clusters comprising many entities, which makes little sense in the context where each cluster should correspond to a single true entity. \cite{betancourt2022random} proposed a framework that overcomes many limitations of \cite{zanella_flexible_2016}, namely a lack of interpretability, identifiability, and full characterization of model asymptotic properties. These papers all proposed subjective priors on the linkage structure that satisfy the microclustering property.

In addition, \cite{betancourt2022prior} proposed subjective priors on the linkage structure using \textit{allelic partitions}. \cite{betancourt2022prior} stated that ``allelic partitions are an
equivalent representation of partitions which summarizes the number of clusters of each size." They considered the microclustering property in a potentially more natural manner. They provided both real and simulation studies illustrating their proposal are similar to previous findings.
All the microclustering papers used the \textit{chaperones algorithm}, which may perform more efficiently than traditional Gibbs sampling \citep{zanella_flexible_2016}.

\paragraph{Takeaways}
We have reviewed entity resolution methods in the literature that may be useful to analysts and tried to point out varying flavors. Supervised methods are the most useful when users have access to training data that is accurate. Unsupervised methods are most useful in situations when users do not have access to unique identifiers, lack confidence in training data, and/or care about uncertainty quantification. The choice of when to use an entity resolution method is application specific. For example, in situations where unique identifiers are not very reliable, 
analysts would prefer unsupervised or semi-supervised methods over fully supervised methods.

\subsection{Canonicalization}
In this section, we review the fourth state of the pipeline, which is \textit{canonicalization}, \textit{data fusion}, or \textit{merging}. This stage attempts to repair any errors from the entity resolution task to create a single integrated data set. The output of this stage is a single integrated data set such that statistical inferences, predictions, and analyses can be performed. It is important to note that each record in the integrated data set is assumed to be unique (with a corresponding unique identifier).

Early methods of canonicalization were rule-based methods as they are easy to implement and scale to large databases \citep{cohen2005incremental}. Much of the literature assumes training data is available to select the most representative record. In many proposals, authors considered decision theory techniques or supervised learning methods \citep{yan1999conflict, bohannon2005cost, culotta2007canonicalization}. See \cite{bleiholder2009data} for a comprehensive review.
In more recent work, \cite{kaplan2022practical} proposed canonicalization methods to choose an integrated (or representative) data set from linked data. The methods are agnostic to the linked method and provide natural uncertainty quantification. They applied their methods to a pipeline approach of blocking, entity resolution, canonicalization, and statistical analysis on a case study of voter registration, illustrating the importance of most of the pipeline approach and its inner workings.

\section{Summary}
The ultimate goal after the data cleaning pipeline is to a perform downstream task or statistical analysis on the integrated data set. 
The goal of this article is to provide an introduction to the entire data cleaning pipeline in a non-technical manner.

The data cleaning pipeline consists of four stages -- schema alignment, blocking, entity resolution, and canonicalization. 
First, schema alignment takes multiple data sources and ensures they are in the same cleaned, tabular format. Second, the aligned schema undergoes blocking, which reduces the dimensionality of the problem. Blocks contain similar records. Third, the analyst performs entity resolution on the blocked data, removing duplications. This stage provides clusters of records that correspond to the same person or object. The fourth stage identifies the most representative record (after entity resolution) to form a single integrated data source (with unique identifiers). This step is essential for predictive and statistical analyses.

\textbf{Acknowledgements}: We thank Kristen Olson, Joseph Sakshaug, and Jenny Thompson for extensive comments and suggestions that immensely improved the quality of the article. We thank Krista Park, Anup Mathur, and MJM for conversations that led to the writing of this article.

\newpage

\bibliographystyle{jasa}
\bibliography{er-review, biblio}

\end{document}